\documentclass[preprint,showpacs,preprintnumbers,amsmath,amssymb]{revtex4}

\usepackage{graphicx}% Include figure files
\usepackage{dcolumn}% Align table columns on decimal point
\usepackage{bm}% bold math

\begin{document}

\title{A Tuneable Few Electron Triple Quantum Dot}% Force line breaks with \\

\author{L. Gaudreau$^{1,2}$, A. Kam$^1$, G. Granger$^1$,  S. A. Studenikin$^1$,  P. Zawadzki$^1$ and A. S. Sachrajda$^1$}
\affiliation{
$^{1}$Institute for Microstructural Sciences, National Research Council, Ottawa, Canada, K1A 0R6\\
$^{2}$Physics Department, University of Sherbrooke, Quebec, Canada, J1K 2R1}

\date{\today}% It is always \today, today,

\begin{abstract}
In this paper we report on a tuneable few electron lateral triple quantum dot design. The quantum dot potentials are arranged in series. The device is aimed at studies of triple quantum dot properties where knowing the exact number of electrons is important as well as quantum information applications involving electron spin qubits. We demonstrate tuning strategies for achieving required resonant conditions such as quadruple points where all three quantum dots are on resonance. We find that in such a device resonant conditions at specific configurations are accompanied by novel charge transfer behaviour.

\end{abstract}

\maketitle

While lateral quantum dots were fabricated since the 1980's, it was only in this decade that such devices were extended to the single electron regime. Our original single few electron quantum dot layout\cite{Ciorga2000} was adapted by Elzerman \textit{et al.} into a double quantum dot potential and included the addition of charge detectors\cite{Elzerman2003}. This classic design led to a series of dramatic experiments in spin manipulation including single electron spin resonance\cite{Koppens2006,Laird2007,Pioro-Ladriere2008} and quantum gates\cite{Petta2005}. Recently, we reported measurements on a few electron triple quantum dot\cite{Gaudreau2006}. The arrangement of dots was incidental in the sense that the location of an underlying impurity played an important role in the definition of the potential topology. Important physics such as quantum cellular automata\cite{Gaudreau2006}, spin blockade\cite{Gaudreau2008} and magneto-conductance effects\cite{Gaudreau2007} were discovered. In this paper we report measurements from a design which creates few electron triple quantum dots in series. Several other groups have recently reported preliminary measurements on triple quantum dot devices\cite{Schroer2007, Rogge2008, Grove-Rasmussen2008, Amaha2008, Amaha2008a}. Few electron triple quantum dot circuits should not be regarded as a stepping stone to a larger system one qubit at a time, but as a device where important concepts can be demonstrated, such as spin buses\cite{Greentree_CTAP} and entanglement\cite{Saraga_Loss} which will certainly be required in more complex architectures for running useful quantum algorithms.

\begin{figure}[h]

\includegraphics[scale=2]{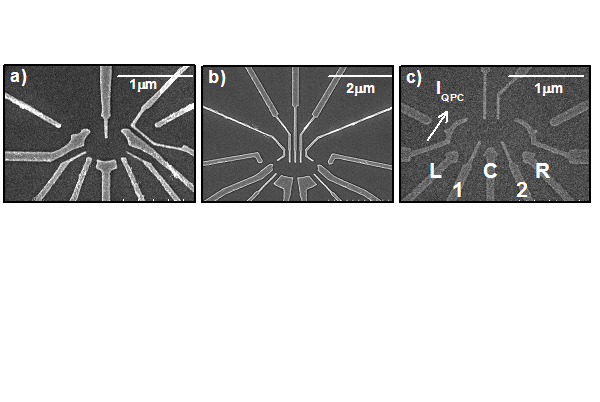}
\caption{SEM images of three different nanostructures designed to create few electron triple quantum dots. As explained in the text, only (c) led to the realization of a \textit{fully tuneable} triple dot system. Relevant gates are labelled as well as the QPC used for charge detection measurements.}
\label{fig:1}
\end{figure}

Figure \ref{fig:1} shows three of the designs we have investigated in the quest for a reliable and tuneable few electron triple quantum dot. Figure \ref{fig:1}(a) is the most direct extension of the original design to triple quantum dots. However, we found that with this layout it was difficult to define an isolated centre dot. The device had a strong tendency to stabilize in a double dot configuration. Figure \ref{fig:1}(b) attempts to circumvent this difficulty by providing gates to divide the potential both from the top and the bottom. While successful in creating multiple quantum dots, it was difficult to generate dots in locations close enough to one another to be useful. In figure \ref{fig:1}(c) we show an optimized combination of the above two devices which we have found to work remarkably well. The data in this paper is from an identical device to that shown in fig. \ref{fig:1}(c).  The approximate location of the three dots is clear from the layout. The two outer quantum dots are made larger to allow the device to have a flexible number of electrons as required. Since the planned experiments did not require the centre dot to hold more than two electrons the centre region is made small. The two top gates as well as two of the bottom gates which protrude into the centre region are used to stabilize the centre quantum dot.

The device was defined on a high mobility GaAs/AlGaAs heterostructure. All of the gates except for gate C were bias cooled with +0.25V to reduce switching noise issues\cite{Pioro-Ladriere2005}. The data presented in this paper was taken using standard low noise AC techniques based on a EG \& G 124A lock-in amplifier and an Ithaco 1211 current preamplifier. Most of the results shown in this paper were measured using one of the Quantum Point Contacts (QPC) as a charge detector. In the usual manner, the QPC was set at a high sensitivity point, where a small change in the electrostatic environment induces a large change in its conductance. A 300$\mu$V bias was applied across the QPC and the current monitored.

\begin{figure}[h]

\includegraphics[scale=2]{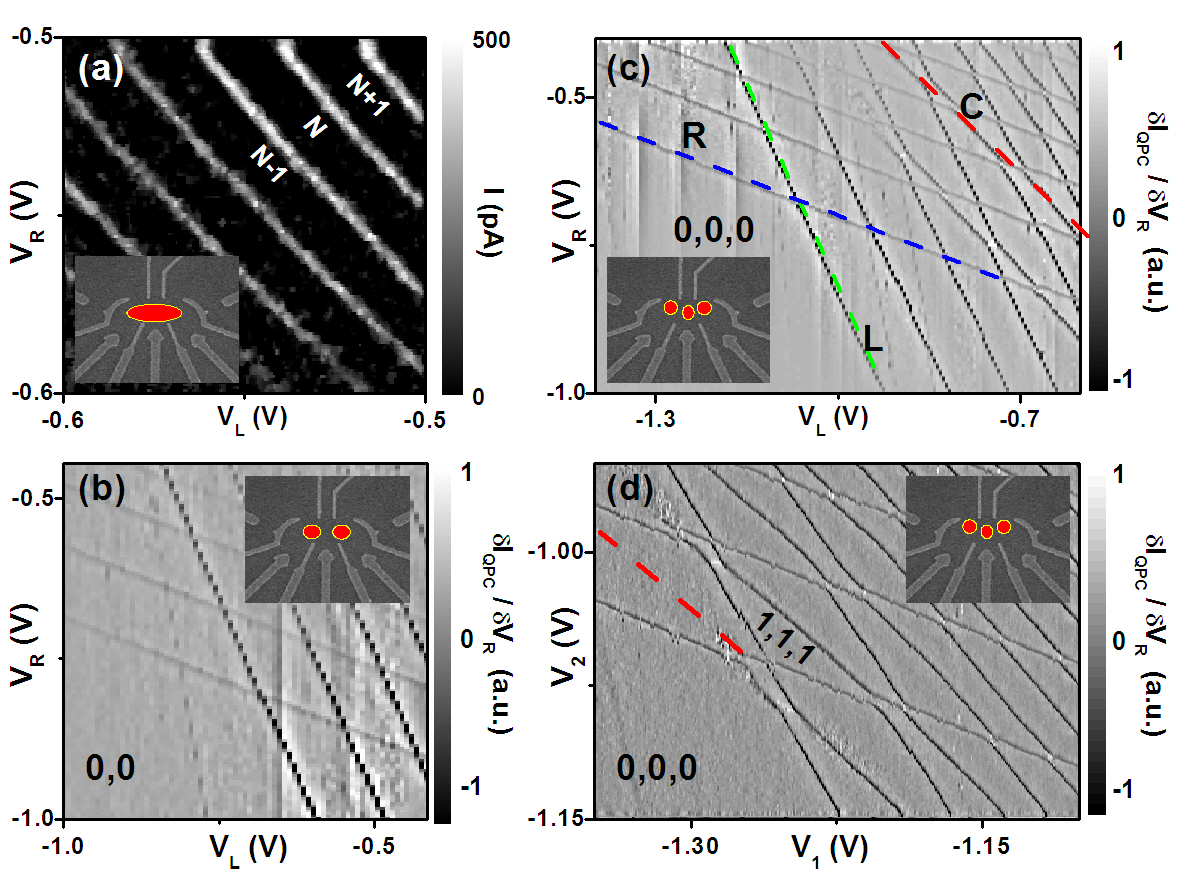}
\caption{Charge stability diagrams of the device used in three different regimes as a function of gate voltages (a) Single dot in the many electron regime. Only one set of parallel lines is observed. (b) Double dot in the few electron regime, where two sets of parallel lines corresponding to the addition spectrum of a dot on the right (nearly horizontal lines) and on the left (nearly vertical lines) sides of the device. (c) Triple dot in the few electron regime. Three sets of parallel lines are visible: green, red and blue corresponding to the left, centre and right dots respectively. (d) Triple dot in the fundamental regime, around the (0,0,0) to (1,1,1) charge configurations. A red dashed line marks the continuation of the addition line for dot C which is not detectable due to pinching of the barriers to the leads. Panel (a) was measured by transport through the single dot, (b), (c), and (d) were measured by transconductance ($\partial I_{QPC}/\partial V_{R}$) of the charge detector. Insets sketch the approximate location of the dots in each configuration.}
\label{fig:2}
\end{figure}

The confining potential in the device shown in fig. \ref{fig:1}(c) can be configured via the applications of different gate voltages into single, double or triple quantum dot modes. Figure \ref{fig:2} shows stability diagrams of this device for a single and a double quantum dot, as well as two settings for the triple dot configuration. The insets show the approximate locations of the dots in each setting. Crossing a line with a particular slope corresponds to adding an electron to a specific dot. In figure \ref{fig:2}(a) (taken using transport techniques) only a single set of parallel lines is visible. In figure \ref{fig:2}(b) two sets of parallel lines exist corresponding to adding electrons to a two dot system. Figure \ref{fig:2}(c) and \ref{fig:2}(d) show two settings for the triple quantum dot. The centre dot contained at most three electrons. An important feature of this design is that it is fairly straightforward to shift the location at which the centre dot exists within the stability diagram. In this way, required few electron configurations can be set up. This capability is shown in fig. \ref{fig:2}(c) and \ref{fig:2}(d) where the third dot exists in different locations. Figure \ref{fig:2}(d) shows the few electron limit of a triple quantum dot.

One of the most crucial tests of a successful triple dot design is the ability to tune it accurately to form quadruple points. A quadruple point, analogous to triple points in double dots, is the location at which all dots are on resonance and four configurations become degenerate. Unlike double quantum dots, the stability diagram of a triple quantum dot is three dimensional, so one is very unlikely to achieve a quadruple point by accident in two dimensional slices of the stability diagram such as fig. \ref{fig:2}(c) and \ref{fig:2}(d). In earlier work\cite{Gaudreau2006}, we developed a strategy for achieving a quadruple point by adjusting relevant gates to shift the two neighbouring triple points on top of one another. We found that for this device an additional tool can be very useful. The two dimensional electron gas (2DEG) regions outside of the quantum dot system can be used for small local side gating adjustments.

\begin{figure}[h]

\includegraphics[scale=2]{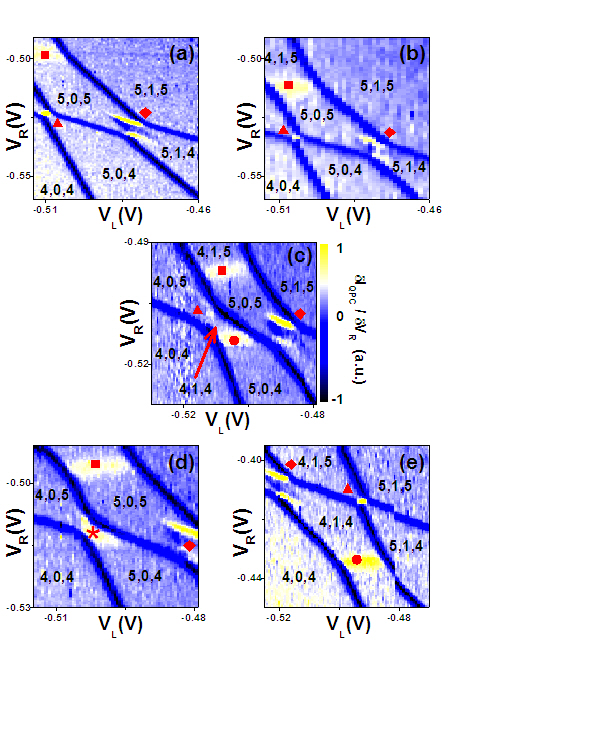}
\caption{Realisation of a quadruple point in which the three dots are on resonance. (a) Three triple points between dots L-C (square), R-C(diamond) and L-R(triangle) are on resonance. (b) Gate voltages are adjusted to bring the three triple points closer to each other. (c) Further gate voltage tuning lead to a QCA effect observed by the appeareance of an extra line (circle) in the vicinity of a quadruple point. (d) Side gating with the 2DEG achieved the exact quadruple point (*) where the circle and the triangle charge transfer lines meet. (e) Gate voltage adjustments lead to the three dots going out of resonance as the three triple points are further apart.}
\label{fig:3}
\end{figure}

Figure \ref{fig:3} reveals how one can accurately set up a quadruple point through a combination of gate voltage and side gating adjustments. In figure \ref{fig:3}(a) one can see three triple point regions. They correspond to resonances between each pair of dots (L-C, R-C, L-R) and the leads. Gate voltages are then fine tuned to adjust the chemical potential of the dots in order to bring them closer to resonance and the stability diagram is retaken and shown in fig. \ref{fig:3}(b). The two triple dot regions are slightly closer together. Gate voltages are further adjusted and fig. \ref{fig:3}(c) is obtained. The two regions are now very close and an extra charge transfer line is visible (marked by a circle). It is parallel to the line marked by a square. This is a signature of a quantum cellular automata (QCA) effect\cite{Gaudreau2006}, which occurs when an electron is added to dot R and as a result an electron from dot C is electrostatically pushed to dot L. To get from 3(c) to the accurate quadruple point, side gating is used. In this specific case this was achieved through the application of an additional 100$\mu$V at the two leads to the dots relative to the 2DEG region on the other side of the left QPC. The quadruple point stability diagram can be seen in fig. \ref{fig:3}(d). The signature of the quadruple point is that two charge transfer lines meet at a single point. This is very clear in fig. \ref{fig:3}(d) and marked by * in the figure. Further voltage adjustments move the two triple point regions through each other. This is shown in fig. \ref{fig:3}(e). Now the  charge transfer line marked by a square disappears as expected since the QCA effect only occurs in the vicinity of the quadruple point.

\begin{figure}[h]

\includegraphics[scale=2]{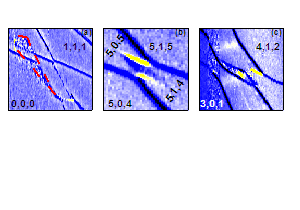}
\caption{Different charging behaviours in the triple dot system for specific charge configurations. (a) Noise region bounded by a triangle in the stability diagram (marked with red dashed lines as a guide to the eye) attributed to backaction from the charge detector. (b) Diamond shaped charge transfer region in a triple point where dots R and C are in resonance. (c) Different charging behaviour in the vicinity of a quadruple point, showing a noise bounded region in the centre.}
\label{fig:4}
\end{figure}

In figure \ref{fig:3} we can see one of several interesting charge transfer characteristics in this device (a diamond shape instead of a charge transfer line). Three of these are shown in fig. \ref{fig:4}. In figure \ref{fig:4}(a) we see a triangular noisy region. Similar triangular regions of noise were previously interpreted as being due to back-action from the QPC in combination with normal bandwidth detection limitations\cite{Taubert2008}. The relaxation processes at the QPC provide the energy source to drive fluctuations between a lead and dot C. In this triangular region the refilling of the electron is slow enough to be picked up in the measurements as the occupation of dot C switches from 0 to 1 electron. Away from these regions the noise disappears because the fluctuations are too fast to be picked up as other channels provide additional rapid relaxation routes\cite{Taubert2008}. The above anomalous charge transfer effect occurs whenever the appropriate tunnel barrier to the lead is pinched off. Figures \ref{fig:4}(b) and \ref{fig:4}(c) reveal charge transfer shapes that have not been observed before and which seem to occur at specific configurations. Figure \ref{fig:4}(b) shows a diamond shape that we observed at certain regions of the stability diagram both between the left and centre dots and remarkably also at some triple points between the left and right dots. The diamond contains no detectable noise within it and no features were observed along the usual charge transfer line connecting the two triple points. However, its presence reveals a more complicated behaviour as one moves from (5,1,4) to (5,0,5). The boundaries correspond in part to extensions of the stability addition lines in the vicinity of triple points. One might expect that this effect is also driven by back-action related to the QPC. Surprisingly, however, a bias dependence measurement (across the QPC) revealed no qualitative effect on the diamond as the bias was increased from 50$\mu$V to 500$\mu$V. Figure \ref{fig:4}(c) shows another charge transfer structure. This one forms as one approaches a specific quadruple point as shown in the figure. The bounded region in this case shows a noisy region within the structure as well as a fairly complex behaviour in the region just outside. It is clear from these structures that the charge transfer processes in a device containing three dots in a line in charge detection measurements is significantly more complex than the corresponding  double dot device. Further experiments on these regions are ongoing and will be published elsewhere.

In summary, we report on a device layout which forms a tuneable few electron triple quantum dot device. We are able to tune it to appropriate configurations, including the fundamental (0,0,0) to (1,1,1) region and place all dots on resonance. In addition, different charge transfer processes are observed at specific configurations in the stability diagram close to locations where two or three dots are on resonance.

We thank P. Hawrylak, M. Korkusinski, J. Kycia, S. Ludwig and M. Pioro-Ladri\`{e}re for useful discussions. ASS acknowledges funding from NSERC. ASS and AK acknowledge funding from CIFAR. We also thank Z. Wasilewski and J. Gupta for the 2DEG wafer.

\newpage

\newpage

\newpage

\newpage

\end{document}